\documentclass[preprint,pra,showpacs,amsmath]{revtex4}
\usepackage{graphicx}
\usepackage{bm}
\begin{document}

\title{Evaluation of Decoherence for
Quantum Control and Computing}

\author{Arkady Fedorov}
\author{Leonid Fedichkin}
\author{Vladimir Privman}
\affiliation{Center for Quantum Device Technology\\
Department of Physics and Department of Electrical and Computer Engineering\\
Clarkson University, Potsdam, NY 13699-5721, USA}

\begin{abstract}
Different approaches in  quantifying environmentally-induced
decoherence are considered. We identify a measure of decoherence,
derived from the density matrix of the system of interest, that quantifies
the environmentally induced error, i.e., deviation from the ideal isolated-system
dynamics. This measure can be
shown to have several useful features. Its behavior as a function of
time has no dependence on the initial conditions, and is expected to be
insensitive to the internal dynamical time scales of the system, thus
only probing the decoherence-related time dependence. For a spin-boson
model---a prototype of a qubit interacting with environment---we
also demonstrate the property of additivity: in the regime of the onset of
decoherence, the sum of the individual qubit error measures
provides an estimate of the error for a several-qubit system, even if the
qubits are entangled, as expected in quantum-computing applications. This
makes it possible to estimate decoherence for several-qubits quantum
computer gate designs for which explicit calculations are exceedingly
difficult.
\end{abstract}


\maketitle

\section{Introduction}

Dynamics of open quantum systems has increasingly attracted \cite{open,CL,Chakravarty,Grabert,vanKampen} the
attention of the community of scientists in diverse fields, working on realizations of quantum information
processing. Recent interest in quantum computing has stimulated
studies \cite{nonMarkov,Anastopoulos,Ford,Braun,Lewis,Wang,Lutz,Khaetskii,OConnell,Strunz,Haake,PMV,short,Privman}
of environmental effects that cause small deviations from the
isolated-system quantum dynamics. To perform large-scale quantum
computation, environment-induced relaxation/decoherence effects
during each short time interval of ``quantum-gate'' functions must
be kept below a certain threshold in order to allow fault-tolerant
quantum error
correction \cite{qec,Steane,Bennett,Calderbank,SteanePRA,Aharonov,Gottesman,Knill}.
The reduced density matrix of the quantum system, with the
environment traced out, is usually evaluated within some
approximation scheme, e.g., \cite{apscheme,SPIE,Zurek,Shnirman}.
In this work, we focus on an \emph{additive\/} measure of the
deviation of the density matrix of a several-qubit system from the ``ideal''
density matrix of the system \cite{norm,dd,addnorm}. The latter would describe
the ``ideal'' dynamics, i.e., the system completely isolated
from the environment. For a spin-boson model of a
single two-state system (a qubit), this measure is calculated
explicitly for the environment modeled as a bath of harmonic
modes \cite{Caldeira}, e.g., phonons or photons. We also establish
that for a several-qubit quantum system, the introduced measure of
decoherence is approximately additive for the time scales of interest,
i.e., for short ``gate function'' times. The fact that this measure
can be evaluated by summing up the deviation measures of
the constituent qubits, which can be, in general, entangled, allows
to avoid lengthy, tedious, and in most cases, intractable, many-body calculations.
This new short-time additivity property is reminiscent of the approximate additivity
expected for relaxation rates of exponential approach to
equilibrium at large times, though the two properties are not related.

Let us briefly outline the commonly accepted scheme for implementing
quantum algorithms in physical systems. The input data are encoded into
the quantum states of several separated two-level systems (qubits). Then,
their evolution is controlled by a Hamiltonian consisting of
single-qubit operators and of two-qubit interaction
terms \cite{Lloyd,Barenco}. Parameters of the Hamiltonian can be
varied (controlled) externally to implement a given algorithm. In most quantum
computer
proposals \cite{Lloyd,Barenco,Lloyd2,Turchette,Cirac,Ekert,Kventsel,Kane,Loss,Imamoglu,Rossi,Nakamura,Tanamoto,Platzman,Sanders,Burkard,Vrijen,Fedichkin,Bandyopadhyay,Larionov} this
control is achieved by changing local electromagnetic fields
around the qubits. Of course, this ideal model does not include the
influence of the environment on the computation, which
necessitates quantum error correction. The latter involves
inevitably the implementation of non-unitary measurement-type
operations \cite{qec,Steane,Bennett,Calderbank,SteanePRA,Aharonov,Gottesman,Knill}
and cannot be described as Hamiltonian-governed dynamics of a closed
system.

To study the effect of the environment, one needs to choose a
suitable model for the environmental and its coupling to the system of interest. The
accepted approach to evaluate environmentally induced decoherence
involves a model in which each qubit is coupled to a bath of
environmental modes \cite{Caldeira}. The reduced density matrix of
the system, with the bath modes traced out, then describes the
time-dependence of the system
evolution \cite{nonMarkov,Anastopoulos,Ford,Braun,Lewis,Wang,Lutz,Khaetskii,OConnell,Strunz,Haake,PMV,Markov,Louisell,Abragam,Blum,apscheme,SPIE,Zurek,Shnirman}.
Due to the interaction with the environment, after each
computational cycle the state of the qubits will deviate slightly
from the ideal state. The deviation
accumulates at each cycle, so that large-scale quantum computation
is not possible without performing fault-tolerant error correction
schemes \cite{qec,Steane,Bennett,Calderbank,SteanePRA,Aharonov,Gottesman,Knill,Kitaev,Kitaev2,Kitaev3}.
These schemes require the environmentally induced decoherence of
the quantum state in one cycle to be below some threshold. The
value of threshold, defined for uncorrelated single qubit error
rates, was estimated \cite{Preskill,DiVincenzo} to be between
$10^{-6}$ and $10^{-4}$.

To study decoherence for a given system, one should first obtain
the evolution of its density matrix. This can be done by using
various approximations \cite{nonMarkov,Anastopoulos,Ford,Braun,Lewis,Wang,Lutz,Khaetskii,OConnell,Strunz,Haake,PMV,short,Markov,Louisell,Abragam,Blum,apscheme,SPIE,Zurek,Shnirman}. During each
computational step the Hamiltonian of the studied system is
usually considered to be constant. The most familiar are the
Markovian-type approximations \cite{Markov,Louisell,Abragam,Blum},
used to evaluate approach to the thermal state at large times. It
has been pointed out recently that these approximations are not
suitable for quantum computing purposes because they are usually
not valid \cite{vanKampen,short} at low temperatures and for
short cycle times of quantum computation. Several non-Markovian
approaches \cite{nonMarkov,Anastopoulos,Ford,Braun,Lewis,Wang,Lutz,Khaetskii,OConnell,Strunz,Haake,PMV,short,Privman,apscheme,SPIE,Zurek}
have been developed to evaluate the short-time dynamics of open
quantum systems.

When one tries to study decoherence of several-qubit systems,
additional difficulties should be taken into account.  Namely, one
has to consider the degree to which noisy environments of
different qubits are correlated \cite{Storcz}. For example, if all
constituent qubits are effectively immersed in the same bath, then
there is a way to reduce decoherence for this group of qubits
without error correction algorithms. The reduction of error rate
can be achieved by encoding the state of one logical qubit in a
decoherence free subspace of the states of several physical
qubits \cite{Palma,DFS,Zanardi,Lidar}.
Therefore, in a large scale quantum information processor
consisting of many thousands of qubits, it is more appropriate to
consider qubits immersed in distinct baths, because these errors
represent the ``worst case scenario'' that necessitates
error-correction.

After obtaining  the density matrix $\rho(t)$ for a single- or
few-qubit system evaluated in some approximation, we have to
compare it to the ideal density matrix $\rho^{(i)}(t)$
corresponding to quantum algorithm without environmental
influences. It is convenient to define some measure of decoherence
to compare with the fault-tolerance
criteria \cite{qec,Steane,Bennett,Calderbank,SteanePRA,Aharonov,Gottesman,Knill}.
It is desirable to have the measure nonnegative, and vanishing if
and only if the system evolves in complete isolation. Since
explicit calculations beyond one or very few qubits are
exceedingly difficult, it would be also useful to find a measure
which is \emph{additive\/} (or at least sub-additive), i.e., the
measure of decoherence, $D$ of a composite system will be the sum
(or not greater than the sum) of the measures of decoherence,  $D_{j}$, of
its subsystems,
\begin{equation}\label{subadditivity}
 D\leq\sum_j D_{j}.
\end{equation}

In this work, we identify a measure of decoherence which has these
desirable properties. In Section \ref{dev}, we give an overview of
different methods for quantifying decoherence. We note that in
some cases these numerical measures have oscillations on the time scales of the
internal system frequencies, which do not reflect the nature of decoherence. Therefore, in the next Section \ref{D(t)} we define the
norm that is subadditive for non-interacting initially unentangled
qubits and in most cases is a monotonic function of time. To
establish stronger properties of subadditivity of this norm even
for initially entangled qubits we will use a more sophisticated
diamond norm. It is defined in Section \ref{K(t)}.
Relation between these norms and conditions for subadditivity
for initially entangled qubits are also discussed. In
Section \ref{specific models}, we consider a specific model of a
qubit interacting with a bosonic bath of environmental modes. For
two types of interaction we explicitly obtain norms $D$ and $K$
for one qubit and prove asymptotic \emph{additivity\/} for short
times, $t$,
\begin{equation}\label{DN4}
 D(t)=\sum_j D_{j}(t)+o\left(\sum_j
D_{j}(t)\right).
\end{equation}
This result should apply as a good approximation up to
intermediate, inverse-system-energy-gap times \cite{short},
and it is consistent with the recent finding \cite{Dalton} that at
short times decoherence of a trapped-ion quantum computer scales
approximately linearly with the number of qubits.

\section{Different Approaches to Quantifying Decoherence} \label{dev}

Typically, the total Hamiltonian of an open quantum system
interacting with environment has the form
\begin{equation}
H =   H_S  +   H_B  + H_I,
\end{equation}
where $H_S$ is the internal system Hamiltonian, $H_B$ is the
Hamiltonian of the environment (bath), $H_I$ is the system-bath
interaction Hamiltonian. Over gate-function cycles, and between
them, the terms in $H$ will be considered constant \cite{norm,dd}.
Therefore, the overall density matrix $R(t)$ of the system and
bath evolves according to
\begin{equation}\label{R(t)}
R(t)=e^{- iHt}R(0)e^{iHt}.
\end{equation}
Here and in the following we use the convention $\hbar=1$.

 Usually the initial density matrix is assumed
\cite{Markov,Louisell,Abragam,Blum} to be a direct product of the
initial density matrix of the system, $\rho(0)$, and the
thermalized density matrix of the bath, $\Theta$,
\begin{equation}R(0)=\rho(0){\raise2pt\hbox{$\scriptscriptstyle{\otimes}$}}
\Theta.\end{equation} The reduced density matrix of the system is obtained by
tracing out the bath modes,
\begin{equation}\label{reduced}
\rho(t)={\rm Tr}_{\, B} \, R(t).
\end{equation}

Usually the environment is assumed to be a large macroscopic
system in thermal equilibrium at temperature $T$. Interaction with
it leads to thermalization of the quantum system, so that the
reduced density matrix at large times approaches
\begin{equation}
\rho \to \frac{{{e^{ - \beta   H_S } }}}{{{{\rm Tr}_{\, S}
\left(e^{ - \beta H_S } \right)}}}, \qquad {\rm as} \;\; t \to
\infty,
\end{equation}\par\noindent
where  $\beta = 1/k_B T$. Markovian type approximations which are
used to quantify this process yield exponential decay of the
density matrix elements in the energy basis of the Hamiltonian
$H_S$ \cite{Markov,Louisell,Abragam,Blum},
\begin{equation}
 \rho _{nn} (t) - \rho _{nn}(\infty) \propto e^{ - t/T_{nn} } ,
\end{equation}
\begin{equation}
 \rho _{nm} (t) \propto e^{ - t/T_{nm} }  \qquad (n \ne m) .
\end{equation}
The shortest times among $T_{nn}$ and $T_{n\neq m}$ are identified
as characteristic times of thermalization $T_1$ and decoherence
$T_2$, respectively. Thermalization of the quantum system requires
the energy exchange with the environment while decoherence can
include other faster processes without energy exchange. Therefore,
it is commonly believed \cite{Blum} that $T_2$ is much shorter than $T_1$
for low-temparature, well isolated from the environment systems appropriate for
quantum computing realizations. The shorter
time $T_2$ can be compared with $T_g$ needed for elementary
quantum gate functions \cite{Privman2}. The ratio $T_g/T_2$ must
be small in order to satisfy fault-tolerance criteria
\cite{Preskill,DiVincenzo}. However, since $T_2$ is a large-time
asymptotic property, other measures, representative of the
short-time, $t \ll T_2$, decoherence properties, are preferred for
actual numerical evaluations \cite{short}.

The exponential behavior of the density matrix elements inherent
to the Markovian approximation is valid
on the time scale which is much larger than the internal
inverse-system-energy-gap times of the quantum system.
To measure effects of decoherence on
these time scales one can use the entropy \cite{Neumann},
\begin{equation}\label{S}
S(t)=- {\rm Tr}\left(  \rho \ln  \rho \right),
\end{equation}\par\noindent
or the idempotency defect, called also the first order
entropy \cite{Kim,Zurek2,Zagur},
\begin{equation} \label{s}
s(t)=1 - {\rm Tr} \left( \rho ^2 \right).
\end{equation}
Both expressions are zero only if the quantum system density
operator is a projector $\rho (0) = | \varphi  \rangle \langle
\varphi |$. Any deviation from a pure state
leads to the increase in the value of both
measures. Expressions (\ref{S}, \ref{s}) provide numerical measure
of the system ``purity'' which does not rely on preferred basis.
However, entropy measures
do not distinguish different pure states.

The next step in analyzing the effect of the interaction with
the environment is to define the ``ideal'' (without interaction)
density operator evolution according to
\begin{equation}\label{ideal}
\rho^{(i)}(t)\equiv e^{-iH_S t}\rho(0)e^{iH_S t}.
\end{equation}
One of the measures that characterizes decoherence in term of the
difference between the ``real'' evolution, $\rho(t)$, and
``ideal'' one, $\rho^{(i)}$, is the fidelity
\cite{Dalton,Fidelity2},
\begin{equation}\label{fidelity}
F(t)={\rm Tr}_{\,S}  \left[ \, \rho^{(i)}(t) \, \rho (t) \,
\right].
\end{equation}
It is particularly useful when $ \rho^{(i)} (t)$
remains a projection operator (pure state) for all times $t \geq
0$, because it then attains the maximum value of 1 only when $\rho (t) = \rho^{(i)} (t)$.

An alternative way to quantify the effect of the interaction with
the bath \cite{norm}, is to consider the deviation, $\sigma(t)$,
\begin{equation}\label{deviation}
\sigma(t)  =   \rho(t)  - \rho^{(i)}(t),
\end{equation}
of the reduced density matrix from the ideal one.
As numerical measures of decoherence one can use the operator
norm $\left\|\sigma \right\|_{\lambda}$ or trace norm
$\left\|\sigma \right\|_{\rm Tr}$. Both norms are standard in
in the theory of linear operators \cite{Kato}. For an
arbitrary linear operator, $A$, these norms are defined as follows,
\begin{equation}\label{n2}
\left\| A \right\|_{\lambda} = \mathop {\sup }\limits_{\varphi \ne
0} \left ( {\frac{{ \langle \varphi |A^ \dagger  A|\varphi \rangle
}}{{ \langle \varphi |\varphi  \rangle }}}\right ) ^{1/2},
\end{equation}
\begin{equation}\label{n2-a}
\left\| A \right\|_{\rm Tr} = {\rm Tr}\,\sqrt { A^{\dagger} A}.
\end{equation}
For a finite-dimensional Hermitian deviation operator (\ref{deviation})
these definitions are equivalent to
\begin{equation}\label{n11}
\left\|\sigma \right\|_{\lambda} = {\max_i} \left| {\lambda _i }
\right|,
\end{equation}
\begin{equation}\label{tracenorm}
\left\|   \sigma  \right\|_{{\rm Tr}}  = \sum\limits_i {\left|
{\lambda _i } \right|},
\end{equation}
where $\lambda_i$ are the eigenvalues of $\sigma$.

In the simplest case of a two-level system (qubit), the two norms are
proportional to each other and given by
\begin{equation}\label{}
\left\| \sigma \right\|_{\lambda} = \sqrt{\left| {\sigma _{11} }
\right|^2 + \left| {\sigma_{12} } \right|^2}={\frac{1}{2}} \left\|
\sigma  \right\|_{{\rm Tr}}.
\end{equation}
The deviation norm $\left\| \sigma
\right\|_{\lambda}$ has the minimal value, 0, only for
$\rho (t) = \rho^{(i)} (t)$ without any additional conditions.

Note that the measures $\left\| \sigma
\right\|_{\lambda}$ and  $\left\| \sigma \right\|_{\rm Tr}$ are not
only functions of time, but also depend on the initial density
operator $\rho(0)$. Due to decoherence, they will deviate from
zero for $t>0$. However, their time-dependence will also contain
oscillations at the frequencies of the internal system dynamics,
as will be illustrated below for $\left\| \sigma
\right\|_{\lambda}$; see \cite{addnorm,norm} and Fig.\ 1. In the
next section we define the \emph{maximal\/} deviation norm,
$D(t)$, which is typically a monotonic function of time.

\section{The Maximal Deviation Norm}\label{D(t)}

To characterize decoherence for an arbitrary initial state, pure
or mixed, we propose to use the maximal norm, $D$, which is
determined as an operator norm maximized over all the possible initial density
matrices. For instance, we can define
\begin{equation}\label{normD}
  D(t) = \sup_{\rho (0)}\bigg(\left\| \sigma (t,\rho (0))\right\|_{\lambda}  \bigg).
\end{equation}

One can show that $0 \leq D(t) \leq 1$.  This measure of
decoherence will typically increase monotonically from zero at
$t=0$, saturating at large times at a value $D(\infty) \leq 1$.
The definition of the maximal decoherence measure $D(t)$ looks
rather complicated for a general multiqubit system. However, we will
show that it can be evaluated in closed form for
short times, appropriate for quantum computing, for a single-qubit
(two-state) system. We then establish an approximate additivity
that allows us to estimate $D(t)$ for several-qubit systems as
well.

In the superoperator notation the evolution of the reduced density
operator of the system (\ref{reduced}) and the one for the ideal
density matrix (\ref{ideal}) can be formally
expressed \cite{Kitaev,Kitaev2,Kitaev3} in the following way
\begin{equation}\label{T1}
\rho(t)=T(t)\rho(0),
\end{equation}
\begin{equation}\label{Ti}
    \rho^{(i)}(t)=T^{(i)}(t)\rho(0),
\end{equation}
where $T$, $T^{(i)}$ are linear superoperators. In this notation
the deviation can be expressed as
\begin{equation}\label{t1}
\sigma(t)=\left[T(t)-T^{(i)}(t)\right]\rho(0).
\end{equation}

The initial density matrix can always be written in the following
form,
\begin{equation}\label{mixture}
    \rho(0)=\sum_{j} p_j |\psi_j\rangle\langle\psi_j|,
\end{equation}
where $\sum_j p_j=1$ and $0 \leq p_j\leq1$. Here the set of the wavefunctions
$|\psi_j\rangle$ is not assumed to have any orthogonality properties.
Then, we get
\begin{equation}\label{}
\sigma\left(t, \rho(0)\right)=\sum_{j} p_j
\left[T(t)-T^{(i)}(t)\right]\left|\psi_j\right\rangle\left\langle\psi_j\right|.
\end{equation}
The deviation norm can thus be bounded,
\begin{equation}\label{proj}
\|\sigma(t, \rho(0))\|_{\lambda} \leq
  \left\|  \left[T(t)-T^{(i)}(t)\right]  |\phi\rangle\langle\phi|\right\|_{\lambda}.
\end{equation}
Here
$|\phi\rangle$ is defined according to
\begin{equation}\nonumber
\left\|  \left[T-T^{(i)}\right]  |\phi\rangle\langle\phi|\right\|_{\lambda}=
\max_j\left\|  \left[T-T^{(i)}\right]  |\psi_j\rangle\langle\psi_j|\right\|_{\lambda}.
\end{equation}
It transpires that for any initial
density operator which is a statistical mixture, one can always find a density operator
which is pure-state,
$|\phi\rangle\langle\phi|$, such that $\|\sigma(t,
\rho(0))\|_{\lambda}\leq\|\sigma(t, |\phi\rangle\langle\phi|)\|_{\lambda}$. Therefore, evaluation of the
supremum over the initial density operators in order to find $D(t)$, see
(\ref{normD}), one can done over only pure-state density operators.

Let us consider strategies of evaluation of $D(t)$ for a single qubit. We can
parameterize $\rho(0)$ as
\begin{equation}\label{parametriazation1}
  \rho(0)=U \left(
\begin{array}{cc}
  P & 0 \\
  0 & 1-P \\
\end{array}
\right)U^{\dagger},
\end{equation}
where $0\leq P \leq 1$, and $U$ is an arbitrary $2 \times 2$
unitary matrix,
\begin{equation}
U=\left(
\begin{array}{cc}
  e^{i(\alpha+\gamma)}\cos\theta & e^{i(\alpha-\gamma)}\sin\theta
\\
  -e^{i(\gamma-\alpha)}\sin\theta & e^{-
i(\alpha+\gamma)}\cos\theta \\
\end{array}\right).
\end{equation}
Then, one should find a supremum of the norm of deviation
(\ref{n11}) over all the possible real parameters $P$, $\alpha$,
$\gamma$ and $\theta$. As shown above, it
suffices to consider the density operator in the form of a
projector and put $P=1$. Thus, one should search for the maximum
over the remaining three real parameters $\alpha$, $\gamma$ and
$\theta$.

Another parametrization of the pure-state density operators,
$\rho(0)=|\phi\rangle\langle\phi|$, is to express an
arbitrary wave function $|\phi\rangle=\sum_j (a_j+i b_j)|j\rangle$
in some convenient ortonormal basis $|j\rangle$, where $j=1,\ldots,N$.
For a two-level system,
\begin{equation}\label{parametrization2}
    \rho(0)=\left(%
\begin{array}{cc}
  a_1^2+b_1^2 &  (a_1+i b_1)(a_2-i b_2) \\
 (a_1-i b_1)(a_2+i b_2) &  a_2^2+b_2^2 \\
\end{array}%
\right),
\end{equation}
where the four real parameters $a_j,b_j$, with $j=1,2$ satisfy
$a_1^2+b_1^2+a_2^2+b_2^2=1$, so that the maximization is again
over three independent real numbers. In the following sections, we
will consider examples of
evaluation of $D(t)$ for single-qubit systems.

In quantum computing, the error rates can be significantly reduced
by using several physical qubits to encode each logical
qubit \cite{DFS,Zanardi,Lidar}. Therefore, even before active
quantum error correction is
incorporated \cite{qec,Steane,Bennett,Calderbank,SteanePRA,Aharonov,Gottesman,Knill},
evaluation of decoherence of several qubits is an important, but
formidable task. Consider a system consisting of two initially
\emph{unentangled\/} subsystems $S_1$ and $S_2$, with decoherence
norms $D_{S_1}$ and $D_{S_2}$, respectively. We denote the density
matrix of the full system as $\rho_{S_1S_2}$ and its deviation as
$\sigma_{S_1S_2}$, and use a similar notation with subscripts
$S_1$ and $S_2$ for the two subsystems. If the evolution of system
 is governed by the ``noninteracting''
Hamiltonian of the form $H_{S_1S_2}=H_{S_1}+H_{S_2}$, where the
terms $H_{S_1}$, $H_{S_2}$ act only on variables of the system
$S_1$, $S_2$, respectively, then the overall norm $D_{S_1S_2}$ can
 be bounded by the sum of the norms $D_{S_1}$ and $D_{S_2}$,
\begin{equation}\label{D12}
D_{S_1S_2}^{\vphantom{(i)}} = \sup_{\rho (0)} \|
\sigma_{S_1S_2}^{\vphantom{(i)}} \|_{\lambda}
  = \sup_{\rho (0)} \| \rho_{S_1S_2}^{\vphantom{(i)}} - \rho^{(i)}_{S_1
S_2}\|_{\lambda}
\end{equation}
\begin{equation}\nonumber
  =\sup_{\rho (0)}\| \rho_{S_1}^{\vphantom{(i)}} \!
{\raise2pt\hbox{$\scriptscriptstyle{\otimes}$}}
\rho_{S_2}^{\vphantom{(i)}} \! - \rho_{S_1}^{(i)}
{\raise2pt\hbox{$\scriptscriptstyle{\otimes}$}}
\rho_{S_2}^{(i)}\|_{\lambda}
  = \sup_{\rho (0)}\| \sigma_{S_1}^{\vphantom{(i)}} \! {\raise2pt\hbox{$\scriptscriptstyle{\otimes}$}}
   \rho_{S_2}^{\vphantom{(i)}} \! + \rho_{S_1}^{(i)} {\raise2pt\hbox{$\scriptscriptstyle{\otimes}$}}
\sigma_{S_2}^{\vphantom{(i)}} \|_{\lambda}
\end{equation}

Since the operator norm obeys the triangle inequality \cite{Kato}
\begin{equation}
\| \sigma_{S_1}^{\vphantom{(i)}} \!
{\raise2pt\hbox{$\scriptscriptstyle{\otimes}$}}
   \rho_{S_2}^{\vphantom{(i)}} \! + \rho_{S_1}^{(i)} {\raise2pt\hbox{$\scriptscriptstyle{\otimes}$}}
\sigma_{S_2}^{\vphantom{(i)}} \|_{\lambda}\leq\|
\sigma_{S_1}^{\vphantom{(i)}} \!
{\raise2pt\hbox{$\scriptscriptstyle{\otimes}$}}
   \rho_{S_2}^{\vphantom{(i)}} \!\|_{\lambda} + \|\rho_{S_1}^{(i)} {\raise2pt\hbox{$\scriptscriptstyle{\otimes}$}}
\sigma_{S_2}^{\vphantom{(i)}} \|_{\lambda},
\end{equation}
we can estimate the last expression as
\begin{equation}\label{D12-a}
  D_{S_1S_2}^{\vphantom{(i)}} =  \sup_{\rho (0)}\| \sigma_{S_1}^{\vphantom{(i)}} \! {\raise2pt\hbox{$\scriptscriptstyle{\otimes}$}}
   \rho_{S_2}^{\vphantom{(i)}} \! + \rho_{S_1}^{(i)} {\raise2pt\hbox{$\scriptscriptstyle{\otimes}$}}
\sigma_{S_2}^{\vphantom{(i)}} \|_{\lambda} \leq \sup_{\rho (0)} \|
\sigma_{S_1}^{\vphantom{(i)}}
{\raise2pt\hbox{$\scriptscriptstyle{\otimes}$}}
\rho_{S_2}^{\vphantom{(i)}} \|_{\lambda}
  +  \sup_{\rho (0)} \|  \rho_{S_1} ^{(i)}
{\raise2pt\hbox{$\scriptscriptstyle{\otimes}$}}
\sigma_{S_2}^{\vphantom{(i)}} \|_{\lambda}.
\end{equation}
Each eigenvalue of the tensor product of two linear operators is
formed as a pairwise product of eigenvalues of the
two operators. Therefore, the operator
norm of the tensor product of two operators is equal to product of
their operator norms,
\begin{equation}\label{A1A2}
 \|A_1{\raise2pt\hbox{$\scriptscriptstyle{\otimes}$}} A_2  \|_{\lambda}= \|A_1  \|_{\lambda} \|A_2  \|_{\lambda}.
\end{equation}
We use this property and the fact that the eigenvalues of density
matrices $\rho_{S_1},\rho_{S_2}$ are in $[0,1]$ to derive the
estimate
\begin{equation}
D_{S_1S_2}^{\vphantom{(i)}}  \leq\sup_{\rho_{S_1} (0)}\|
\sigma_{S_1}^{\vphantom{(i)}} \|_{\lambda} + \sup_{\rho_{S_2}
(0)}\| \sigma_{S_2}^{\vphantom{(i)}}\|_{\lambda} =
D_{S_1}^{\vphantom{(i)}} + D_{S_2}^{\vphantom{(i)}}.
\end{equation}

In general, initially unentangled qubits will remain nearly
unentangled for short times, because they didn't have enough time
to interact. Therefore, the inequality
\begin{equation}\label{Dbound}
 D\lesssim \sum_q D_{q},
\end{equation}
is expected to provide a good approximate estimate for the norm of
a multiqubit system $D$ in terms of the norms $D_q$ calculated in the space of
each individual qubit, i.e., the measures of decoherence for the individual qubits
can be considered \emph{approximately\/} additive. For large
times, the separate measures become of order 1, so such a bound is
not useful. Instead, the \emph{rates\/} of approach of various
quantities to their asymptotic values are approximately additive
in some cases.

In the rest of this work, we focus on the short-time and adiabatic
(i.e., no energy exchange with the bath \cite{basis}) regimes, and
establish a much stronger property: We prove the approximate
additivity for the initially \emph{entangled\/} qubits whose dynamics
is governed by
\begin{equation}
  H=\sum_q H_q=\sum_q \left(H_{Sq}+H_{Bq}+H_{Iq}\right),
\end{equation}
where $H_{Sq}$ is the Hamiltonian of the $q$th qubit itself,
$H_{Bq}$ is the Hamiltonian of the environment of the $q$th qubit,
and $H_{Iq}$ is corresponding qubit-environment interaction. For
this purpose, in the next section we consider a more complicated
(for actual evaluation) diamond norm
\cite{Kitaev,Kitaev2,Kitaev3}, $K(t)$, as an auxiliary quantity
used to establish the additivity of the more easily calculable
operator norm $D(t)$.

\section{The Diamond Norm}\label{K(t)}

The establishment of the upper-bound estimate for the maximal
deviation norm of a multiqubit system, involves several derivations.
We bound this norm by the recently
introduced (in the contexts of quantum computing) \cite{Kitaev,Kitaev2,Kitaev3} diamond norm, $K(t)$.  Actually, for
single qubits, in several models the diamond norm can be
expressed via the corresponding maximal deviation norm. At
the same time, the diamond norm for the whole quantum system is
bounded by sum of the norms of the constituent qubits by using a
specific stability property of the diamond norm.  The use
of  diamond norm was proposed by
Kitaev \cite{Kitaev,Kitaev2,Kitaev3},
\begin{equation}\label{supernormK}
K(t) =\|T- T^{(i)}\|_{\diamond}=\sup_{\varrho} \|
\{[T-T^{(i)}]{\raise2pt\hbox{$\scriptscriptstyle{\otimes}$}} I\}
{\varrho} \|_{\rm Tr}.
\end{equation}
The superoperators $T$, $T^{(i)}$ characterize the actual and
ideal evolutions according to (\ref{T1}, \ref{Ti}), $I$ is the
identity superoperator on a Hilbert space $G$ whose dimension is
the same as that of the corresponding space of the superoperators
$T$ and $T^{(i)}$, and $\varrho$ is an arbitrary density operator
in the product space of twice the number of qubits.

The diamond norm has an important stability property,
proved in \cite{Kitaev,Kitaev2,Kitaev3},
\begin{equation}\label{stability}
\|T_1 {\raise2pt\hbox{$\scriptscriptstyle{\otimes}$}}
T_2\|_{\diamond}=\|T_1\|_{\diamond} \|T_2\|_{\diamond},
\end{equation}
which should  be compared with (\ref{A1A2}). Note that
(\ref{stability}) is a property of the evolution
superoperators rather than that of the density operators: The importance
of this difference will become obvious shortly.

 Consider again a composite system consisting of the two
subsystems $S_1$, $S_2$, with the noninteracting Hamiltonian
\begin{equation}
H_{S_1S_2}=H_{S_1}+H_{S_2}.
\end{equation}
The evolution superoperator of the system will be
\begin{equation}
 T_{S_1S_2}=T_{S_1}{\raise2pt\hbox{$\scriptscriptstyle{\otimes}$}}
T_{S_2},
\end{equation}
and the ideal one
\begin{equation}
T_{S_1S_2}^{(i)}=T_{S_1}^{(i)}{\raise2pt\hbox{$\scriptscriptstyle{\otimes}$}}
T_{S_2}^{(i)}.
\end{equation}
The diamond measure for the system can be expressed as
\begin{eqnarray}
&&K_{S_1S_2}^{\vphantom{(i)}}=\|T_{S_1S_2}^{\vphantom{(i)}} -
T_{S_1S_2}^{(i)}\|_{\diamond}=
\|(T_{S_1}^{\vphantom{(i)}}-T_{S_1}^{(i)}){\raise2pt\hbox{$\scriptscriptstyle{\otimes}$}}
T_{S_2}^{\vphantom{(i)}}+T_{S_1}^{(i)}{\raise2pt\hbox{$\scriptscriptstyle{\otimes}$}}
(T_{S_2}^{\vphantom{(i)}}-T_{S_2}^{(i)})\|_{\diamond}\nonumber\\
&&\leq\|(T_{S_1}^{\vphantom{(i)}}-T_{S_1}^{(i)}){\raise2pt\hbox{$\scriptscriptstyle{\otimes}$}}
T_{S_2}^{\vphantom{(i)}}\|_{\diamond}+\|T_{S_1}^{(i)}{\raise2pt\hbox{$\scriptscriptstyle{\otimes}$}}
(T_{S_2}^{\vphantom{(i)}}-T_{S_2}^{(i)})\|_{\diamond} . \label{justbelow}
\end{eqnarray}
By using the stability property (\ref{stability}), we get
\begin{eqnarray}
&&
K_{S_1S_2}^{\vphantom{(i)}}\leq\|(T_{S_1}^{\vphantom{(i)}}-T_{S_1}^{(i)}){\raise2pt\hbox{$\scriptscriptstyle{\otimes}$}}
T_{S_2}^{\vphantom{(i)}}\|_{\diamond}+\|T_{S_1}^{(i)}{\raise2pt\hbox{$\scriptscriptstyle{\otimes}$}}
(T_{S_2}^{\vphantom{(i)}}-T_{S_2}^{(i)})\|_{\diamond}=
\|T_{S_1}^{\vphantom{(i)}}-T_{S_1}^{(i)}\|_{\diamond}\|
T_{S_2}^{\vphantom{(i)}}\|_{\diamond}
+\nonumber\\
&&\|T_{S_1}^{(i)}\|_{\diamond}\|T_{S_2}^{\vphantom{(i)}}-
T_{S_2}^{(i)}\|_{\diamond}=
\|T_{S_1}^{\vphantom{(i)}}-T_{S_1}^{(i)}\|_{\diamond}+\|T_{S_2}^{\vphantom{(i)}}-
T_{S_2}^{(i)}\|_{\diamond}=K_{S_1}^{\vphantom{(i)}}+K_{S_2}^{\vphantom{(i)}}.
\end{eqnarray}

The approximate inequality
\begin{equation}\label{Kbound}
 K\lesssim \sum_q K_{q},
\end{equation}
for the diamond norm $K(t)$ has thus the same form as for the
norm $D(t)$, (\ref{Dbound}). Let us emphasize that both relations
apply assuming that for short times the subsystem interactions,
directly with each other or via their coupling to the bath modes,
have had no significant effect. However, there is an important
difference in that the relation for $D(t)$ further requires the
subsystems to be initially unentangled. This restriction does not
apply for the relation derived for $K(t)$. This property is particularly
useful for quantum computing, the power of which is based on qubit
entanglement. However, even in the simplest case of the
diamond norm of one qubit, the calculations are extremely
cumbersome. Therefore, the measure $D(t)$ is preferrable for actual
calculations.

The two deviation-operator norms considered Section \ref{dev} are
related by the following inequality
\begin{equation}\label{a1}
\left\|   \sigma  \right\|_{{\lambda}}\leq\frac 1 2 \left\| \sigma
\right\|_{{\rm Tr}}\leq 1.
\end{equation}
Here the left-hand side follows from
\begin{equation}
\rm{Tr} \,\sigma =\sum_j\lambda_j =0.
\end{equation}
It follows
that the $\ell$th eigenvalue of the deviation operator $\sigma$
 that has the maximum absolute value, $\lambda_\ell=\lambda_{\rm{max}}$, can be
expressed as \begin{equation}\lambda_{\ell}=-\sum_{j\neq
\ell}\lambda_j.\end{equation} Therefore, we have
\begin{equation}\label{a2}
 \left\|   \sigma  \right\|_{{\lambda}}=\frac 1 2\left(2 |\lambda_\ell|\right)
 \leq
 \frac 1 2\left(|\lambda_\ell|+\sum_{j\neq \ell}|\lambda_j|\right)=
\frac 1 2\left(\sum_j|\lambda_j|\right)=\frac 1 2\left\|\sigma
\right\|_{\rm Tr}.
\end{equation}
The right-hand side of (\ref{a1}) then also follows, because any density matrix has trace norm 1,
\begin{equation}\label{a3}
\|   \sigma  \|_{{\rm Tr}} = \|   \rho-\rho^{(i)} \|_{{\rm
Tr}}\leq \|   \rho \|_{{\rm Tr}}+ \|\rho^{(i)} \|_{{\rm Tr}}=2.
\end{equation}

From the relation (\ref{a3}) it follows that
\begin{equation}\label{prop}
 K(t)\le 2.
\end{equation}
 By taking supremum of the both sides of the relation (\ref{a2}) we get
\begin{equation}\label{prop1}
 D(t)=\sup_{\rho(0)}\left\|   \sigma  \right\|_{{\lambda}}\le \frac 12 \sup_{\rho(0)}\left\|   \sigma  \right\|_{\rm Tr}
 \le\frac 12 K(t),
\end{equation}
where the last step involves technical derivation details not reproduced here.
In fact, in the following sections we show that for a single qubit, calculations
within selected models actually give
\begin{equation}\label{}
 D(t)=\frac 12 K(t).
\end{equation}
Since $D$ is generally bounded by (or equal to) $K/2$,
it follows that the multiqubit norm $D$ is
approximately bounded from above by the sum of the single-qubit
norms even for the \emph{initially entangled\/} qubits,
\begin{equation}\label{DN1}
    D(t) \le \frac 12 K(t) \lesssim \frac 12 \sum_q K_{q}= \sum_q D_{q},
\end{equation}
where $q$ labels the qubits.

\section{Decoherence in the Short-Time Approximation} \label{specific models}

Typically the environment, a large macroscopic system, is modelled
by a bath of an infinite number of modes. Each mode is
represented by its own Hamiltonian $M_k$,
\begin{equation}\label{HB}
H_B  = \sum\limits_k  \,M_k.
\end{equation}
The interaction with the bath is often described by the coupling of its modes to
 Hermitian operator $\Lambda_S$ of the quantum system,
\begin{equation}\label{HI}
H_I  = \Lambda _S \sum\limits_k  \,J_k.
\end{equation}
For a bosonic-mode heat bath \cite{Leggett}  we take
\begin{equation}
M_k= \omega_k a_k^{\dagger}
 a_k^{\vphantom{\dagger}},\quad J_k=g_k^{\vphantom{\dagger}} a_k^{\dagger}+
g_k^{*\vphantom{\dagger}} a_k^{\vphantom{\dagger}}.
\end{equation}
Here $\omega_k$ are the bath mode frequencies,
$a_k^{\vphantom{\dagger}},\, a_k^{\dagger}$ are the bosonic
annihilation and creation operators, and $g_k$ are the coupling
constants. Two eigenbases of the operators $H_S$ and $\Lambda_S$
are
\begin{equation}\label{energy basis}
H_S |n\,\rangle  = E_n |n\,\rangle ,\quad \Lambda_S
|\gamma\,\rangle =\lambda_{\gamma} |\gamma\,\rangle.
\end{equation}

At the the initial time $t=0$ the total density matrix of the
system and bath is a direct product
$R(0)=\rho(0){\raise2pt\hbox{$\scriptscriptstyle{\otimes}$}}\Theta$
of the initial density matrix of the system $\rho(0)$ and the
density matrix of the bath $\Theta$. The latter is a product
$\Theta=\theta_1{\raise2pt\hbox{$\scriptscriptstyle{\otimes}$}}\,\theta_2\cdots$
of the bath modes density matrices $\theta_k$. Each bath mode $k$
is assumed to be thermalized,
\begin{equation}\label{mode density operator}
\theta_k = \frac{e^{ - \beta M_k }} {\mathop
{\,{\rm{Tr}}\,}\nolimits_k \left(\,e^{ - \beta M_k } \right)}.
\end{equation}

In the short-time approximation \cite{short} the exponentials in
(\ref{R(t)}) representing the time evolution of the total density
matrix $R(t)$ are approximated as
\begin{equation}\label{expon}
e^{i(H_S  + H_B  + H_I )t + O(t^3 )}  = e^{iH_S t/2} \,e^{i(H_B  +
H_I )t} \,e^{iH_S t/2}.
\end{equation}
The matrix elements of the reduced density matrix $\rho(t)$ in the
free energy basis can be expressed as
\begin{eqnarray}\label{rho_mn}
\rho _{mn} (t) &=& \mathop {\,{\rm{Tr}}\,}\nolimits_B \langle
m|e^{ - iH_S t/2} \,e^{ - i(H_B  + H_I )t} \,e^{ - iH_S t/2}
R(0)\,e^{iH_S t/2} \,e^{i(H_B  + H_I )t} \,e^{iH_S t/2} |n\rangle\
\end{eqnarray}
After cumbersome calculations \cite{short}, utilizing the fact
that the trace over the bath modes can be carried out separately
for each mode, for the bosonic bath case one obtains
\begin{equation}\label{short-time1}
\rho _{mn} (t) = \sum\limits_{p, q, \mu , \nu } \langle m|\mu
\rangle\langle \mu |p\rangle\langle q|\nu \rangle\langle \nu |n
\rangle e^{i[ ( E_q + E_n  - E_p  - E_m )t/2]} \rho _{pq} (0) e^{
- B^2 (t)(\lambda _\mu - \lambda _\nu )^2 /4 + i
C(t)(\lambda_\mu^2  - \lambda_\nu^2 )},
\end{equation}where
\begin{equation}\label{spectral function}
 B^2 (t) \equiv 8 \sum\limits_k {} \displaystyle \frac{{\left | {g_k }
\right | ^2 }}{{\omega _k ^2 }}\sin ^2 {\displaystyle{\frac{\omega
_k t}{2}}}\coth {\displaystyle{\frac{\beta \omega _k }{2}}},
\end{equation}
\begin{equation}
 C(t) \equiv \sum\limits_k {} \displaystyle\frac{{\left| {g_k }
\right|^2 }}{{\omega _k ^2 }}(\omega _k t - \sin \omega _k
 t).
\end{equation}
Here the Roman-labeled states, $|i\rangle$, are the eigenstates of
$ H_S$ corresponding to the eigenvalues $E_i$, with $i = m,n,p,q$.
The Greek-labeled states, $|\zeta \rangle$, are the eigenstates of
$\Lambda_S$ with the eigenvalues $\lambda_\zeta $, where $\zeta =
\mu, \nu$. Details, more general expressions, and additional
discussion can be found in \cite{short}.

\section{The Spin-Boson Model}

Let us consider a system which is a spin-1/2 particle in an
applied magnetic field, interacting with the boson bath. In this
case the system Hamiltonian is
\begin{equation}
H_S={-\frac{\Omega}{2}}\sigma_z
\end{equation}
and interaction  can be chosen as  $\Lambda_S=\sigma_x$, where
$\sigma_x$ and $\sigma_z$ are the Pauli matrices, and $\Omega>0$
is the energy gap between the ground (up, $|1\rangle=\mid\uparrow
\rangle$) and excited (down, $|2\rangle=\mid\downarrow \rangle$)
states of the qubit. The eigenstates of $ \sigma_x$ will be
denoted by
\begin{equation}\label{4}
   \sigma _x | \pm \rangle  = \pm | \pm \rangle ,
\end{equation}\par\noindent
where
\begin{equation}
    |\pm\rangle= \frac{1}{{\sqrt 2}}\left(|1\rangle  \pm |2
\rangle\right).
\end{equation}

The dynamics of the system can be obtained in closed form as
\begin{equation}\label{short-time}
\rho _{mn} (t)=\sum\limits_{{p, q=1,2} \atop {\mu , \nu = \pm }}
\langle m|\mu \rangle\langle \mu |p\rangle\langle q|\nu
\rangle\langle \nu |n \rangle \rho _{pq} (0) e^{i(n-m)\delta_{mp}
\delta_{nq}\Omega t- B^2 (t)(1-\delta_{\mu\nu})},
\end{equation}
where $\delta_{mn}$ is Kronecker symbol.

Note that this result depends only on the spectral function
$B^2(t)$, defined in (\ref{spectral function}), because
\begin{equation}
\lambda _\mu ^2 = \lambda _\nu  ^2=1.
\end{equation}
This function is obtained by integration over the bath mode
frequencies. When the summation in (\ref{spectral function}) is
converted to integration in the limit of infinite number of the
bath modes \cite{basis,vanKampen,Palma}, we get
\begin{equation} \label{Bint}
 B^2 (t) = 8 \int d \omega N(\omega) |g(\omega)|^2 \omega^{-2}
\sin ^2 {\displaystyle{\frac{\omega t} {2}}}
 \coth {\displaystyle{\frac{\beta \omega} {2}}} ,
\end{equation}\par\noindent
where $N(\omega)$ is the density of states. In many realistic
models of the bath, the density of states increases as a power of
$\omega$ for small frequencies and has a cutoff, $\omega_c$, at large
frequencies (Debye cutoff in the case of a phonon bath).
Therefore, approximately setting
\begin{equation}  \label{Bint2}
 N(\omega) |g(\omega)|^2  \propto \omega^n
 \exp \left(- \omega / \omega_c \right)
\end{equation}\par\noindent
can yield a good qualitative estimate of the relaxation
behavior \cite{vanKampen,Palma}. For a popular case of Ohmic
dissipation \cite{Leggett}, $n=1$ and the function $B^2(t)$ has the
initial stage of quadratic growth, intermediate region of
logarithmic growth, and linear-in-$t$ large-time behavior.

Evaluation of
 (\ref{short-time}) yields the following expressions,
\begin{equation}\label{evolution}
\rho _{ 22 } (t) = \left [ 1 + e^{ - B^2 (t)} \right ]\frac{\rho
_{ 22 } (0)}{2} + \left[1 - e^{ - B^2 (t)}\right]\frac{\rho _{ 11
} (0)}{2},
\end{equation}
\begin{equation}\label{evolution0}
\rho _{ 21 } (t) = e^{ -i\Omega t} \left[1 + e^{ - B^2 (t)}
\right]\frac{\rho _{ 21} (0)}{2} + \left[1 - e^{ - B^2 (t)}
\right]\frac{\rho _{ 12 } (0)}{2}.
\end{equation}
Deviation operator $\sigma\left(t\right)$ is defined by
\begin{equation}\label{evolution1}
\sigma _{22}(t)=\frac{1}{2}\left[1-e^{-B^2(t)}\right]\left[\rho
_{11}(0)-\rho _{22}(0)\right],
\end{equation}
\begin{equation}\label{evolution1-1}
\sigma _{21}(t)=\frac{1}{2}\left[1-e^{-B^2(t)}\right]\left[\rho
_{12}(0)-e^{-i\Omega t}\rho _{21}(0)\right].
\end{equation}
With $\rho_{12}(0)=|\rho_{12}(0)|e^{i \phi}$, we get
\begin{eqnarray}\label{n1qubit}
\left\| \sigma(t)  \right\|_{\lambda}&=& \hbox{$\frac{1}{2}$}\big[
1 - e^{ - B^2 (t)}\big] \big\{ [\rho _{11} (0) - \rho _{22} (0)]^2
  +  4 \left| \rho _{12} (0) \right|^2 \sin ^2 [(\Omega
/2)t +\phi] \big\} ^{1/2}.
\end{eqnarray}
In Fig.\ 1, we show schematically the behavior of $ \left\|
\sigma(t)  \right\|_{\lambda} $  for three representative choices
of the initial density matrix $ \rho (0)$. Generally, the norm $ \left\|
\sigma(t)  \right\|_{\lambda} $ increases with time, reflecting
the decoherence of the system. However, oscillations at the
system's internal frequency $\Omega$ are superimposed, as seen
explicitly in (\ref{n1qubit}). Thus, the decohering effect of
the bath is better quantified by the maximal
operator norm, $D(t)$. Explicit calculations yield the result, shown in
Fig.\ 1,
\begin{equation}\label{D1qubit}
D (t)  =\frac{1}{2} \left[ 1 - e^{ - B^2 (t)}\right],
\end{equation} which is indeed a monotonically increasing function of time.
\begin{figure}
\includegraphics[width=16cm, height=13cm]
{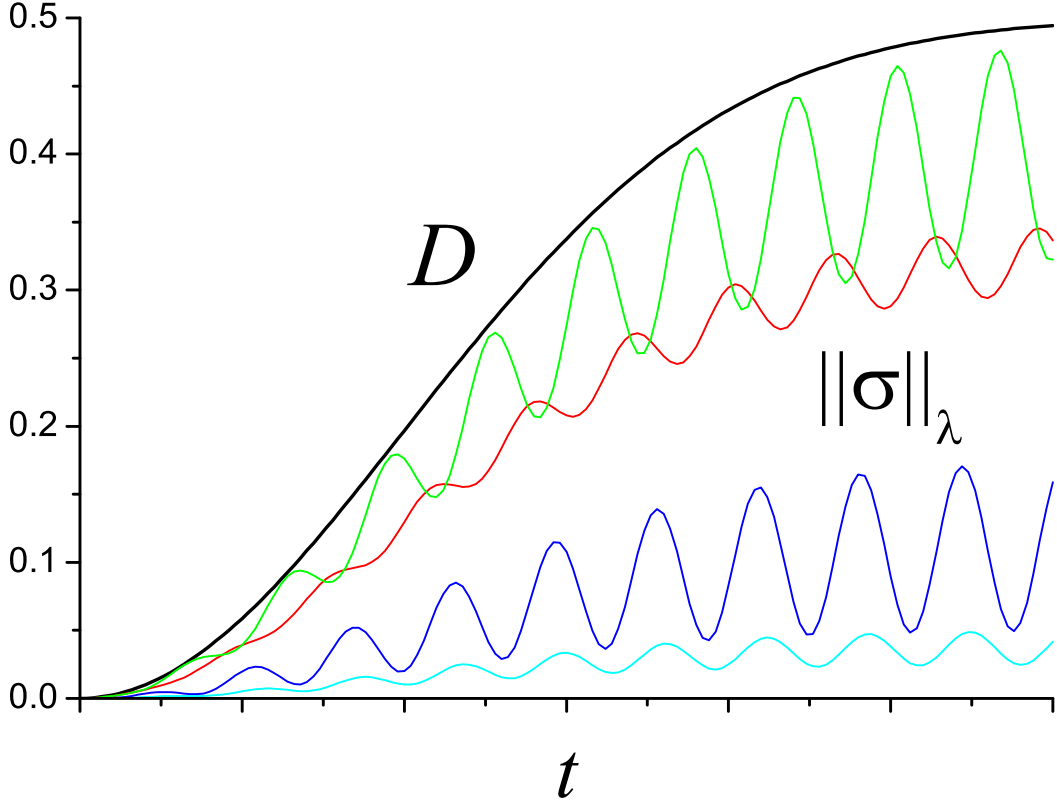} \caption{The maximal deviation norm $D$ vs. time for
the spin-boson model with Ohmic bath. The non-monotonic curves illustrate
the behaviour of the norm $ \left\|
\sigma(t)  \right\|_{\lambda} $ for
several initial choices of density operator $\rho(0)$.}
\end{figure}

Let us now consider the diamond norm $K(t)$ for one qubit. We will
later use the result to estimate the norm $D(t)$ for the
multi-qubit case. To find $K(t)$ for a two-level system one has to
deal with the $4\times 4$ density matrix $\varrho_{jk, lm}$, where
$j,k,l,m=1,2$. To evaluate
\begin{equation}
\varSigma\equiv\{\left(T-T^{(i)}\right){\raise2pt\hbox{$\scriptscriptstyle{\otimes}$}}
I\}\varrho,
\end{equation}
one assumes that $T-T^{(i)}$ acts in the subspace labeled by the
indices $j,l$, see (\ref{evolution1},\ref{evolution1-1}), while
the subspace labeled by the remaining pair of indexes, $k,m$, is
unaffected. The resulting expression are
\begin{equation}\label{varsigma}
\varSigma_{2k, 2m}=\frac{1}{2}\left[ 1 - e^{ - B^2 (t)}\right]
\left[\varrho _{1k,1m}   - \varrho _{2k,2m}  \right],
\end{equation}
\begin{equation}\nonumber
\varSigma_{2k,1m}=\frac{1}{2}\left[1-e^{-B^2(t)}\right]\left[\varrho
_{1k,2m} -e^{-i\Omega t}\varrho _{2k,1m} \right],
\end{equation}
\begin{equation}\nonumber
\varSigma_{1k, 1m}=-\varSigma_{2k, 2m},\quad
\varSigma_{1k,2m}=\varSigma_{2m,1k}^*,\quad k,m=1,2.
\end{equation}
The maximal trace norm of the $4\times4$ matrix
$\varSigma$ is calculated by considering the pure-state density matrices $\varrho=|\phi\rangle\langle
\phi|$, similarly to the consideration in Section
\ref{D(t)}, see (\ref{t1}-\ref{parametrization2}). Here
$|\phi\rangle$ can be expressed in
terms of the basis states,
\begin{equation}\label{phi}
|\phi\rangle=\sum_{m,n=1,2} (a_{nm}+ib_{nm}) |nm\rangle.
\end{equation}
The constants $a_{nm}$, $b_{nm}$ are normalized real amplitudes, such
that $\sum_{m,n=1,2} (a_{nm}^2+b_{nm}^2)=1$. The eigenvalues
of $\varSigma$, see (\ref{varsigma}), are $\varsigma _{1,2}=0$ and
\begin{widetext}
\begin{eqnarray}\label{Kst}
\varsigma_{3,4}=\pm\left[\frac{1-e^{-B^2(t)}}{2}\right]
\left\{1-4\left[(a_{11}a_{21}+a_{12}a_{22}+b_{11}b_{21}+b_{12}b_{22})\cos
\left(\frac{\Omega t}{2}\right) \right.\right.&&\nonumber\\
-\left.\left.(a_{21}b_{11}+a_{22}b_{12}-a_{11}b_{21}-a_{12}b_{22})\sin
\left(\frac{\Omega t}{2}\right)\right]^2\right\}^{1/2}.
\end{eqnarray}
\end{widetext}
The maximal value corresponds to the square root in the expression
(\ref{Kst}) equal to 1, and thus the diamond norm is
\begin{equation}\label{Kshort}
K(t)= 1-e^{-B^2(t)}.
\end{equation}

The above calculation establishes that $K(t)=2D(t)$ for the spin-boson model at short times, and (\ref{DN1}) gives
the upper bound on the multiqubit norm $D$,
\begin{equation}
    D \lesssim\sum_q D_{q}=\frac 1 2 \sum_q \left[1-e^{-B_q^2(t)}\right].
\end{equation}

For short times, one can also establish a lower bound on $D(t)$.
Consider a specific initial state with all the $Q$ qubits excited,
$\rho(0)=(|2\rangle\langle 2|)_1
{\raise2pt\hbox{$\scriptscriptstyle{\otimes}$}} \ldots
{\raise2pt\hbox{$\scriptscriptstyle{\otimes}$}} (|2\rangle\langle
2|)_Q$. Then according to (\ref{evolution},\ref{evolution0}),
$\rho(t) = \rho_1 (t)
{{\raise2pt\hbox{$\scriptscriptstyle{\otimes}$}}} \ldots
{\raise2pt\hbox{$\scriptscriptstyle{\otimes}$}} \rho_Q(t)$, where
\begin{equation}\label{}
\rho_q (t)=\frac{1}{2} \left\{\left[1-e^{- B^2_q(t)}\right]|1
\rangle\langle 1|+ \left[1+e^{-B^2_q(t)}\right]|2 \rangle\langle
2|\right\}.
\end{equation}
The right-bottom matrix element of the (diagonal) deviation
operator,
\begin{equation}\label{}
\sigma_{2^Q,2^Q}(t)=-1+2^{-Q}{\prod_q}\left[1+e^{-B^2_q(t)}\right],
\end{equation}
can be expanded, for small times, as
\begin{equation}
 \frac 12 \sum_q B_q^2 (t)+o\left(\sum_q B_q^2 (t)\right),
\end{equation}
because $B(t)$ vanishes for $t \to 0$. The largest eigenvalue of
$\sigma (t)$ cannot be smaller than $\sigma_{2^Q,2^Q}(t)$. It
follows that
\begin{equation}\label{DN0}
  D\geq \frac 1 2 \sum_q B^2_q
+o\left(\sum_q B_q^2 \right)= \sum_q D_{q}+o\left(\sum_q
D_{q}\right),
\end{equation}
where we used (\ref{D1qubit}) for short times.

By combining the
upper and lower bounds, we get the final result for short times,
\begin{equation}\label{DN4-a}
    D(t)=\sum_q D_{q}(t)+o\left(\sum_q
D_{q}(q)\right).
\end{equation}

\section{Spin Model for Pure Decoherence}

Let us consider a two-level system interacting with the bosonic
bath of the environmental modes adiabatically, i.e., without
exchange of energy with the bath modes \cite{basis}. While energy
exchange processes are needed for thermalization, and also
contribute to decoherence, additional, ``pure'' decoherence
processes are possible and are expected to be important at low
temperatures and short to intermediate times, appropriate for
quantum computing designs \cite{short}. The spin-model
Hamiltonians $H_S$ and $H_B$ will still be assumed of the form
\begin{equation}
H_S={-\frac{\Omega}{2}}\sigma_z,
\end{equation}
\begin{equation}
H_B=\sum\limits_k \omega_k a_k^{\dagger}a_k.
\end{equation}
However, the interaction term will be now of the form commuting with the
system's energy, thus making the latter conserved,
\begin{equation}\label{interadiab}
H_I=\sigma _z \sum_k {\left(g_k^{\vphantom{\dagger}}
a_k^{\dagger}+ g_k^{*\vphantom{\dagger}} a_k^{\vphantom{\dagger}}
\right)}.
\end{equation}

Since $H_S$ commutes with $H_I$, instead of the
approximate formula (\ref{expon}) we have the exact factorization,
\begin{equation}\label{exactexpon}
e^{i(H_S  + H_B  + H_I )t}  = e^{iH_S t/2} \,e^{i(H_B  + H_I )t}
\,e^{iH_S t/2}.
\end{equation}
Therefore, the analytical expression for the reduced density
operator (\ref{short-time1}) is exact and valid for all
times \cite{vanKampen, basis}. For the two-level case \cite{Palma},
\begin{equation}\label{mm}
\rho_{11}(t) = \rho _{11} (0),
\end{equation}
\begin{equation}\label{mm-a}
\rho_{22}(t) = \rho _{22} (0),
\end{equation}
\begin{equation}\label{mmm}
\rho _{12} (t)=\rho _{12} (0)e^{i\Omega t - B^2( t)},
\end{equation}
where the spectral function $B(t)$ was defined in
(\ref{spectral function}). The diagonal density matrix elements
remain constant because there is no energy exchange of the system
with the bath. However, there is pure (adiabatic) decoherence manifested by the decay
of the off-diagonal elements, characterized by the spectral
function $B(t)$ (\ref{mmm}). The corresponding deviation matrix is expressed as follows,
\begin{equation}\label{sigma0}
\sigma_{11}(t) =\sigma_{22}(t)=0,
\end{equation}
\begin{equation}
\sigma_{12}=\rho _{12} (0)e^{i\Omega t- B^2( t)}.
\end{equation}

 The operator norm of $\sigma$ is
\begin{equation}\label{sigma1}
  \left\|\sigma(t) \right\|_{\lambda}=\left[1-e^{ - B^2( t)}\right]\left|\rho _{12}
  \right|.
\end{equation}
Since both diagonal elements of the density matrix and its
eigenvalues are in $[0,1]$ it follows that the absolute value of
the off-diagonal element of any two-level-system density matrix
less than $1/2$,
\begin{equation}
\left|\rho _{12} \right| \leq \frac 12.
\end{equation}
Thus,
\begin{equation}\label{Dadiab}
 D(t)=\sup_{\rho (0)}\big(\| \sigma (t,\rho (0))\|_{\lambda}
  \big) = \frac{1}{2}\left[ 1 - e^{ - B^2
 ( t)}\right].
\end{equation}

One can also evaluate
\begin{equation}
 \varSigma=\{\left(T-T^{(i)}\right){\raise2pt\hbox{$\scriptscriptstyle{\otimes}$}}
I\}\varrho,
\end{equation}
\begin{equation}
 \varSigma_{jk,jm}=0,
 \end{equation}
\begin{equation}
\varSigma_{1k,2m}=\varSigma_{2m,1k}^*=\varrho_{1k,2m}
(0)e^{i\Omega t- B^2
 ( t)}
\end{equation}
For $\varrho=|\phi\rangle\langle \phi|$, with $|\phi\rangle$ as in
(\ref{phi}), the eigenvalues of $\varSigma$ are
$\varsigma_1=\varsigma_2=0$ and
\begin{equation}
\varsigma_{3,4}=\pm \left[1 - e^{ -
B^2(t)}\right]\sqrt{\left(\varrho_{11,11}+\varrho_{12,
12}\right)\left(\varrho_{21, 21}+\varrho_{22, 22}\right)}.
\end{equation}
Since
\begin{equation}
\rm{Tr}\varrho=1,
\end{equation}
one can show that
\begin{equation}
\lambda_{3,4}=\pm \left[1 - e^{ - B^2(t)}\right]\sqrt{x(1-x)},
\end{equation}
with
\begin{equation}
x=\varrho_{11, 11}+\varrho_{12, 12}
\end{equation}
satisfying $0\leq x\leq 1$. The
 diamond norm (\ref{supernormK}) thus follows,
\begin{equation}\label{Kadiab}
K(t) = \underset{ {0\leq x \leq 1}}{\sup}\left\{2\left[1 - e^{ -
B^2(t)}\right]\sqrt{x(1-x)}\right\}= 1 - e^{ - B^2(t)}.
\end{equation}

It is instructive to compare (\ref{Dadiab},\ref{Kadiab}) and
(\ref{D1qubit},\ref{Kshort}). The results are identical despite
the fact that the interaction terms and
density matrix time-dependence
(\ref{evolution},\ref{evolution0},\ref{mm},\ref{mmm}) are different. As in the case of the short-time approximation,
we get the upper bound for the multiqubit norm $D(t)$, (\ref{DN1}).

To establish the lower bound on $D(t)$, we consider a specific initial state with
$\rho(0)=|\Psi\rangle\langle\Psi|$, which is a superposition of
the state corresponding to all the qubits in their ground states
and that of all qubits in their excited states,
\begin{equation}\label{}
|\Psi\rangle=\frac {1}{\sqrt 2} \left(|1 \ldots 1\rangle+|2 \ldots
2\rangle\right).
\end{equation}
Then according to (\ref{mm},\ref{mmm}),
\begin{equation}\label{}
\rho_{1,1}(t)=\rho_{2^Q,2^Q}(t)=\frac{1}{2},
\end{equation}
\begin{equation}
\rho_{1,2^Q}(t)=\frac 1 2\exp{\Big[-i\sum_q \Omega_q t-\sum_q
B^2_q(t)\Big]}.
\end{equation}
The only non-zero matrix elements of the deviation operator are
the right-top and left-bottom matrix elements,
\begin{equation}
\sigma_{1,2^Q}(t)=-\frac{1}{2} \left[1-e^{-\sum_q
B^2_q(t)}\right]e^{i t \sum_q\Omega_q},
\end{equation}
For short times, the absolute value of $\sigma_{1,2^Q}(t)$  can be expressed as
\begin{equation}
\frac{1}{2}\sum_q B_q^2 (t)+o\left(\sum_q B_q^2 (t)\right),
\end{equation}
where the first term gives the largest
eigenvalue of $\sigma (t)$. It follows that
\begin{equation}\label{DN0-a}
  D\geq \frac{1}{2}\sum_q B^2_q
+o\left(\sum_q B_q^2 \right)= \sum_q D_{q}+o\left(\sum_q
D_{q}\right),
\end{equation}
where we used (\ref{Dadiab}) for short times. Finally, we get the
same result (\ref{DN4-a}) for the approximate additivity of $D$ for
short times, for the present model of adiabatic decoherence,
\begin{equation}\label{DN4-b}
    D(t)=\sum_q D_{q}(t)+o\left(\sum_q
D_{q}(t)\right).
\end{equation}

In summary, we introduced the maximal operator norm
suitable for evaluation of decoherence for quantum system immersed
in a noisy environment. The new maximal operator norm
was evaluated for spin models with two types of
bosonic bath interaction. We established both general and model specific subadditivity and additivity properties of this measure of decoherence for
multi-qubit system at short times. The latter property allows evaluation of decoherence
for complex systems in the regime of interest for quantum computing applications.

This research was supported by the National Security Agency
and Advanced Research and Development Activity under Army Research
Office contract DAAD-19-02-1-0035, and by the
National Science Foundation, grant DMR-0121146.


\begin{thebibliography}{15}
\bibitem{open}
G. W. Ford, M. Kac and P. Mazur, J. Math. Phys. \textbf{6}, 504
(1965).

\bibitem{CL}  A. O. Caldeira and A. J. Leggett, Physica A
\textbf{121}, 587 (1983).

\bibitem{Chakravarty} S. Chakravarty and A. J. Leggett, Phys. Rev. Lett.
\textbf{52}, 5 (1984).

\bibitem{Grabert}  H. Grabert, P. Schramm and G.-L. Ingold, Phys. Rep. \textbf{168}, 115 (1988).

\bibitem{vanKampen}
N. G. van Kampen, J. Stat. Phys. \textbf{78}, 299 (1995).

\bibitem{nonMarkov}
K. M. Fonseca Romero and M. C. Nemes, Phys. Lett. A \textbf{235},
432 (1997).

\bibitem{Anastopoulos} C. Anastopoulos and B. L. Hu, Phys.
Rev. A \textbf{62}, 033821 (2000).

\bibitem{Ford} G. W. Ford and R. F. O'Connell, Phys. Rev. D \textbf{64},
105020 (2001).

\bibitem{Braun} D. Braun, F. Haake and W. T. Strunz, Phys. Rev. Lett.
\textbf{86}, 2913 (2001).

\bibitem{Lewis} G. W. Ford, J. T. Lewis and R. F. O'Connell, Phys. Rev. A
\textbf{64}, 032101 (2001).

\bibitem{Wang} J. Wang, H. E. Ruda and B. Qiao, Phys. Lett. A \textbf{294}, 6
(2002).

\bibitem{Lutz} E. Lutz, Phys. Rev. A {\bf 67}, 022109 (2003); cond-mat/0208503 (2002).

\bibitem{Khaetskii} A. Khaetskii, D. Loss and L. Glazman, Phys. Rev. B
\textbf{67}, 195329 (2003).

\bibitem{OConnell} R. F. O'Connell and J. Zuo, Phys. Rev. A \textbf{67}, 062107
(2003).

\bibitem{Strunz} W. T. Strunz, F. Haake and D. Braun, Phys. Rev. A \textbf{67},
022101 (2003).

\bibitem{Haake} W. T. Strunz and F. Haake, Phys. Rev. A \textbf{67}, 022102
(2003).

\bibitem{PMV} V. Privman, D. Mozyrsky and I. D. Vagner, Comp. Phys.
Commun. \textbf{146}, 331 (2002).

\bibitem{short}
V. Privman, J. Stat. Phys. \textbf{110}, 957 (2003).

\bibitem{Privman}
V. Privman, Mod. Phys. Lett. B \textbf{16}, 459 (2002).


\bibitem{qec} P. W. Shor, Phys. Rev. A {\bf 52},  R2493 (1995).

\bibitem{Steane} A. M. Steane, Phys. Rev. Lett. {\bf 77}, 793 (1996).

\bibitem{Bennett} C. H. Bennett, G. Brassard, S. Popescu, B. Schumacher,
J. A. Smolin and W. K. Wootters, Phys. Rev. Lett. {\bf 76}, 722
(1996).

\bibitem{Calderbank} A. R. Calderbank and P. W. Shor, Phys. Rev. A {\bf 54}, 1098
(1996).

\bibitem{SteanePRA} A. M. Steane, Phys. Rev. A {\bf 54}, 4741 (1996).

\bibitem{Aharonov} D. Aharonov and M. Ben-Or, quant-ph/9611025 (1996).

\bibitem{Gottesman} D. Gottesman, Phys. Rev. A {\bf 54}, 1862
(1997).

\bibitem{Knill} E. Knill and R. Laflamme, Phys. Rev. A {\bf 55}, 900 (1997).

\bibitem{apscheme}
D. Loss and D. P. DiVincenzo, cond-mat/0304118 (2003).

\bibitem{SPIE} V. Privman, Proc. SPIE \textbf{5115}, 345 (2003).

\bibitem{Zurek} W. H. Zurek, Rev. Mod. Phys. \textbf{75}, 715 (2003).

\bibitem{Shnirman}
A. Shnirman and G. Sch\"on, cond-mat/0210023 (2002).

\bibitem{norm}
L. Fedichkin, A. Fedorov and V. Privman, Proc. SPIE \textbf{5105},
243 (2003).

\bibitem{dd}
L. Fedichkin  and A. Fedorov, quant-ph/0309024 (2003).

\bibitem{addnorm}
L. Fedichkin, A. Fedorov and V. Privman, cond-mat/0309685 (2003).

\bibitem{Caldeira}
A. O. Caldeira and A. J. Leggett, Phys. Rev. Lett. \textbf{46},
211 (1981).

\bibitem{Lloyd} S. Lloyd, Phys. Rev. Lett. \textbf{75}, 346
(1995).

\bibitem{Barenco}
A. Barenco, C. H. Bennett, R. Cleve, D. P. DiVincenzo,
N. Margolus, P. Shor, T. Sleator, J. A. Smolin and H. Weinfurter,
Phys. Rev. A \textbf{52}, 3457 (1995).

\bibitem{Lloyd2}
S. Lloyd, Science {\bf 261}, 1569 (1993).

\bibitem{Turchette}
Q. A. Turchette, C. J. Hood, W. Lange, H. Mabuchi, and
H. J. Kimble, Phys. Rev. Lett. {\bf 75}, 4710 (1995).

\bibitem{Cirac}
J. I. Cirac and P. Zoller, Phys. Rev. Lett. {\bf 74}, 4091 (1995).

\bibitem{Ekert}
A. Ekert and R. Jozsa, Rev. Mod. Phys. {\bf 68}, 733 (1996).

\bibitem{Kventsel} V. Privman,
I. D. Vagner and G. Kventsel, Phys. Lett. {\bf A239}, 141 (1998).

\bibitem{Kane} B. E. Kane, Nature {\bf 393}, 133 (1998).

\bibitem{Loss} D. Loss and D. P. DiVincenzo, Phys. Rev. A {\bf 57}, 120
(1998).

\bibitem{Imamoglu} A. Imamoglu, D. D. Awschalom, G. Burkard, D. P. DiVincenzo,
D. Loss, M. Sherwin and A. Small, Phys. Rev. Lett. {\bf 83}, 4204
(1999).

\bibitem{Rossi}
P. Zanardi and F. Rossi, Phys. Rev. B {\bf 59}, 8170 (1999).

\bibitem{Nakamura}
Y. Nakamura, Yu. A. Pashkin, and H. S. Tsai, Nature {\bf 398}, 786
(1999).

\bibitem{Tanamoto}
T. Tanamoto, Phys. Rev. A {\bf 61}, 022305 (2000).

\bibitem{Platzman}
P. M. Platzman and M. I. Dykman, Science {\bf 284}, 1967 (1999).

\bibitem{Sanders}
G. P. Sanders, K. W. Kim, W. C. Holton, Phys. Rev. A {\bf 60},
4146 (1999); quant-ph/9909070 (1999).

\bibitem{Burkard}
G. Burkard, D. Loss, D. P. DiVincenzo, Phys. Rev. B {\bf 59}, 2070
(1999).

\bibitem{Vrijen}
R. Vrijen, E. Yablonovitch, K. Wang, H. W. Jiang, A. Balandin,
V. Roychowdhury, T. Mor and D. P. DiVincenzo, Phys. Rev. A {\bf
62}, 012306 (2000).

\bibitem{Fedichkin}
L. Fedichkin, M. Yanchenko and K. A. Valiev, Nanotechnology {\bf
11}, 387 (2000).

\bibitem{Bandyopadhyay} S. Bandyopadhyay, Phys. Rev. B {\bf 61}, 13813 (2000).

\bibitem{Larionov}
A. A. Larionov, L. E. Fedichkin, and K. A. Valiev, Nanotechnology
{\bf 12}, 536 (2001).

\bibitem{Markov}
N. G. van Kampen, \emph{Stochastic Processes in Physics and
Chemistry\/}, North-Holland, 2001.

\bibitem{Louisell} W. H. Louisell, \emph{Quantum Statistical Properties of
Radiation\/}, Wiley, 1973.

\bibitem{Abragam} A. Abragam, \emph{The Principles of Nuclear Magnetism\/},
Clarendon Press, 1983.

\bibitem{Blum}
K. Blum, \emph{Density Matrix Theory and Applications\/}, Plenum,
1996.

\bibitem{Kitaev}
A. Y. Kitaev, Russ. Math. Surv. \textbf{52}, 1191 (1997).

\bibitem{Kitaev2}
 D. Aharonov, A. Kitaev and N. Nisan, Proc. XXXth ACM
Symp. Theor. Comp., Dallas, TX, USA, 20 (1998).

\bibitem{Kitaev3}
A. Yu. Kitaev, A. H. Shen and M. N. Vyalyi, \emph{Classical and
Quantum Computation\/}, AMS, 2002.

\bibitem{Preskill}
J. Preskill, Proc. Roy. Soc. Lond. A {\bf 454}, 385 (1998).

\bibitem{DiVincenzo} D. P. DiVincenzo, Fort. Phys. \textbf{48}, 771
(2000).

\bibitem{Storcz}
M. J. Storcz and F. K. Wilhelm, Phys. Rev. A \textbf{67}, 042319
(2003).

\bibitem{Palma}
G. M. Palma, K. A. Suominen and A. K. Ekert, Proc. Roy. Soc. Lond.
A \textbf{452}, 567 (1996).

\bibitem{DFS}
L.-M. Duan and G.-C. Guo, Phys. Rev. Lett. \textbf{79}, 1953
(1997).

\bibitem{Zanardi} P. Zanardi and M. Rasetti, Phys. Rev. Lett. \textbf{79},
3306 (1997).

\bibitem{Lidar} D. A. Lidar, I. L. Chuang and K. B. Whaley, Phys. Rev. Lett. \textbf{81}, 2594 (1998).

\bibitem{Dalton}
B. J. Dalton, J. Mod. Opt. \textbf{50}, 951 (2003).

\bibitem{Privman2}
V. Privman, D. Mozyrsky and I. D. Vagner, Comp. Phys. Commun.
\textbf{146}, 331 (2002).


\bibitem{Neumann}
J. von Neumann, \emph{Mathematical Foundations of Quantum
Mechanics\/}, Princeton University Press, 1983.


\bibitem{Kim}
J. I. Kim, M. C. Nemes, A. F. R. de Toledo Piza and H. E. Borges,
  Phys. Rev. Lett. \textbf{77}, 207 (1996).

\bibitem{Zurek2}
W. H. Zurek, S. Habib and J. P. Paz,
  Phys. Rev. Lett. \textbf{70}, 1187 (1993).

\bibitem{Zagur}
J. C. Retamal and N. Zagury,
 Phys. Rev. A \textbf{63}, 032106 (2001).

\bibitem{Fidelity2}
L.-M. Duan and G.-C. Guo, Phys. Rev. A \textbf{56}, 4466 (1997).

\bibitem{Kato}
T. Kato, \emph{Perturbation Theory for Linear Operators\/},
Springer-Verlag, 1995.

\bibitem{Leggett}
A. J. Leggett, S. Chakravarty, A. T. Dorsey, M. P. A. Fisher, A.
Garg and W. Zwerger, Rev. Mod. Phys. \textbf{59}, 1 (1987).

\bibitem{basis}
D. Mozyrsky and V. Privman, J. Stat. Phys. \textbf{91}, 787
(1998).

\end{thebibliography}
\end{document}